\documentclass[]{aa}
\bibpunct{(}{)}{;}{a}{}{,}

\usepackage[colorlinks]{hyperref}
\hypersetup{allcolors={blue}}

\usepackage[varg]{txfonts}
\usepackage{relsize,exscale}
\usepackage{color}

\title{SN~2018ijp: the explosion of a stripped-envelope star within a dense H-rich shell?\thanks{Photometric tables are only available at the CDS via anonymous ftp.}\fnmsep\thanks{Spectroscopic data are available through the Weizmann Interactive Supernova Data Repository (WISeREP).}
}

\newcommand{\ha}{H$\alpha$}
\newcommand{\hb}{H$\beta$}
\newcommand{\hg}{H$\gamma$}
\newcommand{\hd}{H$\delta$}
\newcommand{\kms}{\,km\,s$^{-1}$}
\newcommand{\msun}{\,M$_{\sun}$}
\newcommand{\ang}{\,\AA}
\newcommand{\ergs}{\,erg\,s$^{-1}$}
\renewcommand{\object}{SN~2018ijp}

\makeatletter
\renewcommand*\aa@pageof{, page \thepage{} of \pageref*{LastPage}}
\makeatother

\begin{document}
\author{L.~Tartaglia\inst{1,2}\href{https://orcid.org/0000-0003-3433-1492}
\and{J.~Sollerman}\inst{1}\href{https://orcid.org/0000-0003-1546-6615}
\and{C.~Barbarino}\inst{1}\href{https://orcid.org/0000-0002-3821-6144}
\and{F.~Taddia}\inst{1}\href{https://orcid.org/0000-0002-2387-6801}
\and{E.~Mason}\inst{3}\href{https://orcid.org/0000-0003-3877-0484}
\and{M.~Berton}\inst{4,5}\href{https://orcid.org/0000-0002-1058-9109}
\and{K.~Taggart}\inst{6}\href{https://orcid.org/0000-0002-5748-4558}
\and{E.~C.~Bellm}\inst{7}\href{https://orcid.org/0000-0001-8018-5348}
\and{K.~De}\inst{8}\href{https://orcid.org/0000-0002-8989-0542}
\and{S.~Frederick}\inst{9}\href{https://orcid.org/0000-0001-9676-730X}
\and{C.~Fremling}\inst{10}\href{https://orcid.org/0000-0002-4223-103X}
\and{A.~Gal-Yam}\inst{11}\href{https://orcid.org/0000-0002-3653-5598}
\and{V.~Z.~Golkhou}\inst{7,12}\href{https://orcid.org/0000-0001-8205-2506}
\and{M.~Graham}\inst{10}\href{https://orcid.org/0000-0002-3168-0139}
\and{A.~Y.~Q. Ho}\inst{8}\href{https://orcid.org/0000-0002-9017-3567}
\and{T.~Hung}\inst{13}\href{https://orcid.org/0000-0002-9878-7889}
\and{S.~Kaye}\inst{14}
\and{Y.-L.~Kim}\inst{15}\href{https://orcid.org/0000-0002-1031-0796}
\and{R.~R.~Laher}\inst{16}\href{https://orcid.org/0000-0003-2451-5482}
\and{F.~J.~Masci}\inst{16}\href{https://orcid.org/0000-0002-8532-9395}
\and{D.~A.~Perley}\inst{6}\href{https://orcid.org/0000-0001-8472-1996}
\and{M.~D.~Porter}\inst{14}
\and{D.~J.~Reiley}\inst{14}
\and{R.~Riddle}\inst{14}\href{https://orcid.org/0000-0002-0387-370X}
\and{B.~Rusholme}\inst{16}\href{https://orcid.org/0000-0001-7648-4142}
\and{M.~T.~Soumagnac}\inst{17,11}\href{https://orcid.org/0000-0001-6753-1488}
\and{R.~Walters}\inst{10,14}
}
\institute{
Department of Astronomy and the Oskar Klein Centre, Stockholm University, AlbaNova, SE 106 91 Stockholm, Sweden \\(\email{leonardo.tartaglia@astro.su.se})
\and{INAF - Osservatorio Astronomico di Padova, Vicolo dell’Osservatorio 5, I-35122 Padova, Italy}
\and{INAF - Osservatorio Astronomico di Trieste, Via G.B.~Tiepolo, 11, I-34143 Trieste, Italy}
\and{Finnish Centre for Astronomy with ESO (FINCA), University of Turku, Vesilinnantie 5, FI-20014 University of Turku, Finland}
\and{Aalto University Mets{\"a}hovi Radio Observatory, Mets{\"a}hovintie 114, FI-02540 Kylm{\"a}l{\"a}, Finland}
\and{Astrophysics Research Institute, Liverpool John Moores University, IC2, Liverpool Science Park, 146 Brownlow Hill, Liverpool L3 5RF, UK}
\and{DIRAC Institute, Department of Astronomy, University of Washington, 3910 15th Avenue NE, Seattle, WA 98195, USA}
\and{Cahill Center for Astrophysics, California Institute of Technology, MC 249-17, 1200 E California Boulevard, Pasadena, CA 91125, USA}
\and{Department of Astronomy, University of Maryland, College Park, MD 20742, USA}
\and{Division of Physics, Mathematics and Astronomy, California Institute of Technology, Pasadena, CA 91125, USA}
\and{Department of Particle Physics and Astrophysics, Weizmann Institute of Science, 234 Herzl St., Rehovot, 76100, Israel}
\and{The eScience Institute, University of Washington, Seattle, WA 98195, USA}
\and{Department of Astronomy and Astrophysics,University of California, Santa Cruz, California, 95064, USA}
\and{Caltech Optical Observatories, California Institute of Technology, Pasadena, CA  91125  USA}
\and{Universit\'e de Lyon, Universit\'e Claude Bernard Lyon 1, CNRS/IN2P3, IP2I Lyon, F-69622, Villeurbanne, France}
\and{IPAC, California Institute of Technology, 1200 E. California Blvd, Pasadena, CA 91125, USA}
\and{Lawrence Berkeley National Laboratory, 1 Cyclotron Road, Berkeley, CA 94720, USA}
}
\date{Submitted to A\&A on 2020 August 3, first review submitted on 2021 February 21, Accepted on 2021 April 15}

\abstract{In this paper, we discuss the outcomes of the follow-up campaign of SN~2018ijp, discovered as part of  the Zwicky Transient Facility survey for optical transients.
Its first spectrum shows similarities to broad-lined Type Ic supernovae around maximum light, whereas later spectra display strong signatures of interaction between rapidly expanding ejecta and a dense H-rich circumstellar medium, coinciding with a second peak in the photometric evolution of the transient.
This evolution, along with the results of modeling of the first light curve peak, suggests a scenario where a stripped star exploded within a dense circumstellar medium.
The two main phases in the evolution of the transient could be interpreted as a first phase dominated by radioactive decays, and an later interaction-dominated phase where the ejecta collide with a pre-existing shell.
We therefore discuss SN~2018jp within the context of a massive star depleted of its outer layers exploding within a dense H-rich circumstellar medium.}


\titlerunning{A SN~Ic-BL interacting with its circumstellar medium}
\authorrunning{L.~Tartaglia et al.}

\maketitle

\section{Introduction} \label{sec:intro}
A supernova (SN) is the most spectacular way a star can end its life, where progenitors more massive than $8-9$\msun~\citep[see, e.g.,][]{2003ApJ...591..288H,2009ARA&A..47...63S} are expected to explode as core-collapse (CC) SNe.

SNe interacting with a dense circumstellar medium (CSM) can produce a wide range of observables, resulting in a large heterogeneity of photometric and/or spectroscopic properties.
The classification of interacting transients is typically based on the presence of narrow emission features in their spectra, with Type IIn \citep{1990MNRAS.244..269S} or Ibn \citep{2016MNRAS.456..853P,2019ApJ...871L...9H} SNe being those showing prominent narrow H or He lines, respectively.

The current picture for the most common narrow-lined interacting SNe is that of fast moving ejecta colliding with a slow-moving dense CSM.
In the shocked regions, a characteristic ``forward-reverse" shock structure forms \citep[e.g.,][]{1982ApJ...258..790C,1994ApJ...420..268C}, and energetic photons can ionize the surrounding medium giving rise to the structured, multi-component emission line profiles usually observed in SNe IIn \citep[see, e.g.,][]{1993MNRAS.262..128T,1994MNRAS.268..173C,2020A&A...638A..92T}.
In this context, narrow emission lines (full--width--at--half--maximum -- FWHM -- of a few $10^2$\kms) are recombination lines produced in the slow-moving, un-shocked CSM.

This requires the presence of a dense CSM produced by the progenitor star prior to its explosion, and seems to suggest massive luminous blue variables \citep[LBVs, see, e.g.,][]{2008A&A...483L..47T,2009Natur.458..865G}, red supergiants (RSG) with super-winds \citep[see, e.g.,][]{2009AJ....137.3558S,2010ApJ...717L..62Y} or Wolf-Rayet (WR) stars in binary systems \citep[e.g.,][]{2016ApJ...833..128M} as candidate progenitors for SNe IIn and Ibn.
Such stars are all able to provide the environment required to produce the signatures of interaction \citep[see][]{1994PASP..106.1025H,2007ARA&A..45..177C}.
This scenario is also supported by observations of eruptive episodes occurring weeks to years prior to the explosion of the star as an interacting SN \citep{2007ApJ...657L.105F,2007Natur.447..829P,2013ApJ...767....1P,2014ApJ...789..104O,2016MNRAS.459.1039T}.

On the other hand, ejecta-CSM interaction can occur in many kinds of explosions or stellar outbursts and may prevent the observer to gain insight on the intrinsic nature of the transient, including the explosion mechanism triggering the SN explosion.
This is the case for the sub-class of interacting transients known as SNe Ia-CSM \citep[see, e.g.,][]{2013ApJS..207....3S}, which are believed to be thermonuclear explosions embedded in a dense H-rich medium. 
\begin{figure}
\begin{center}
\includegraphics[width=0.95\columnwidth]{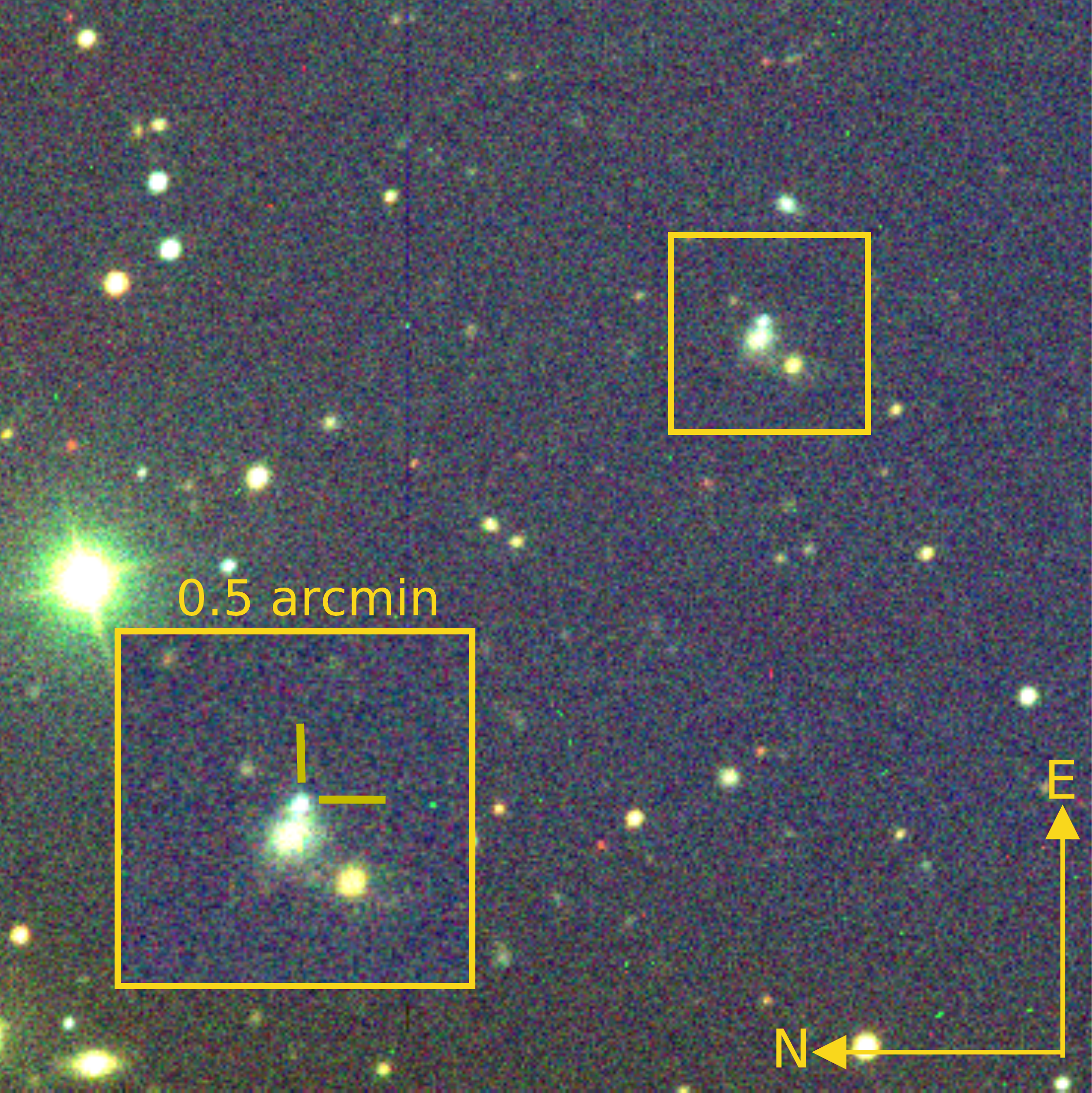}
\caption{Color image of the field of \object, obtained combining $g-$, $r-$ and $i-$band frames obtained on 2019 January 26 with LT. \object~is the the blue source in the middle of the inset. \label{fig:rgb}}
\end{center}
\end{figure}

Signatures of interaction, occasionally delayed with respect to the SN explosion, have been observed also in stripped-envelope (SE) SNe, optical transients typically showing a lack of H (SNe IIb and Ib) or both H and He features (SNe Ic) in their optical spectra.
A few recent examples of SE SNe showing this evolution are SNe~2014C \citep{2015ApJ...815..120M,2017ApJ...835..140M}, 2017dio \citep{2018ApJ...854L..14K}, 2017ens \citep{2018ApJ...867L..31C} and SNe~2019tsf and 2019oys \citep{2020A&A...643A..79S}. These can be interpreted as CC explosions of SE SNe within a dense, H-rich medium, with a possible progenitor scenario being that of a massive CSM produced during the RSG stage or by a stellar companion \citep[e.g.,][]{2015ApJ...815..120M}.

In this context, we present the discovery and discuss the results of our follow-up campaign of \object, discovered as part of the Zwicky Transient Facility \citep[ZTF;][]{2019PASP..131g8001G,2019PASP..131a8002B,2019PASP..131a8003M} during the first year of operations.
The transient was discovered in the host SDSS~J102137.72+085554.1 (Fig.~\ref{fig:rgb}) on 2018 November 7.41~UT and labelled ZTF18aceqrrs\footnote{\url{https://lasair.roe.ac.uk/object/ZTF18aceqrrs/}}. 
The photometric and spectroscopic follow-up campaigns were triggered soon after discovery through the GROWTH `Marshal' \citep{2019PASP..131c8003K}.
A description of the facilities used and the reduction steps performed to obtain final light curves and spectra are described in Sect.~\ref{sec:dataredu}.
\object~shows a relatively fast photometric evolution with double-peaked $g-$ and $r-$band light curves. 
The spectra show strong signatures of delayed interaction with a dense pre-existing H-rich CSM in the form of narrow H lines in emission increasing their strength with time and a spectral continuum becoming significantly bluer with time, as described in Sect.~\ref{sec:analysis}.
While interaction features dominate the evolution of \object~at later times, we note that the spectrum around the first peak resembles those typical of a subclass of broad-lined Type Ic SNe \citep[Ic-BL SNe; see, e.g.,][and references therein]{2019A&A...621A..71T}, with a good match with several Type Ic/Ic-BL SNe.
In addition, modeling the first peak in the context of a radioactively powered light curve gives $^{56}\rm{Ni}$ and total ejected masses comparable with those obtained for the Type Ic-BL SN iPTF15dqg \citep{2019A&A...621A..71T}.
In Sect.~\ref{sec:analysis} and \ref{sec:classification}, we therefore discuss the observables of \object~in the context of a massive, SE star within a dense H-rich medium.
A brief summary of our main conclusions is reported in Sect.~\ref{sec:conclusions}.

In the following, we adopt a foreground Galactic extinction $E(B-V)=0.029\,\rm{mag}$ along the line of sight of \object, as estimated by \citet{2011ApJ...737..103S} using a standard extinction law \citep{1989ApJ...345..245C} with $R_V=3.1$.
We did not include any additional contribution from the local environment to the total extinction, since we could not identify strong \ion{Na}{ID} features at the redshift of the host in the spectra of \object~(see Sect.~\ref{sec:spectroscopy}).
The distance to \object~was computed from the redshift derived using host lines (see Sect.~\ref{sec:spectroscopy}) assuming a standard cosmology with $\rm{H_0}=73\,\rm{km}\,\rm{s^{-1}}\,\rm{Mpc^{-1}}$, $\Omega_M=0.27$ and $\Omega_{\Lambda}=0.73$, resulting in a luminosity distance $D_L=373\,\rm{Mpc}$\footnote{Derived using {\sc CosmoCalc} \citep{2006PASP..118.1711W} available at: \url{http://www.astro.ucla.edu/~wright/CosmoCalc.html}}.
\begin{table*}
\caption{Log of the spectroscopic observations of SN~2018ijp}
\label{table:18ijpspeclog}
\resizebox{\linewidth}{!}{\begin{tabular}{cccccccc}
\hline\hline
Date & JD & Phase & Instrumental setup & Grism/Grating & Spectral range & Resolution & Exposure time \\
& & (d) & & & (\AA) & ($\lambda/\Delta\lambda$) & (s) \\
\hline
20181201 & 2458454.08 & $+24$  & Keck1+LRIS   & 400/3400+400/8500   & $3500-9500$  & 900              & $300+300$     \\
20190115 & 2458498.64 & $+65$  & NOT+ALFOSC   & Gr4                 & $4000-9500$  & 300              & 2700          \\
20190201 & 2458516.03 & $+81$  & Keck1+LRIS   & 400/3400+400/8500   & $3500-9500$  & 860              & $600+600$     \\
20190204 & 2458519.58 & $+84$  & NOT+ALFOSC   & Gr4                 & $4000-9500$  & 320              & 2700          \\
20190227 & 2458542.44 & $+105$ & NOT+ALFOSC   & Gr4                 & $4000-9500$  & 400              & $2\times2700$ \\
20190403 & 2458576.89 & $+137$ & Keck1+LRIS   & $400/3400+400/8500$ & $3500-9500$  & 800              & $300+300$     \\
20190504 & 2458607.56 & $+165$ & VLT+Xshooter & UVB+VIS+NIR         & $3500-20000$ & $5400+8900+5600$ & $3\times(1200+1262+300)$ \\
20200124 & 2458873.02 & $+410$ & Keck1+LRIS   & $400/3400+400/8500$ & $3500-9500$  & 850              & $1375$        \\

\hline
\end{tabular}}
\tablefoot{The resolution of each spectrum was estimated from \ion{[O}{I]} sky lines. Rest frame phases refer to the estimated epoch of the explosion. NOT: $2.56\,\rm{m}$ Nordic Optical Telescope with ALFOSC; VLT: $8\,\rm{m}$ Very Large Telescope with X-shooter (ESO Observatorio del Paranal, Chile); KECK: $10\,\rm{m}$ Keck I telescope with LRIS (Mauna Kea Observatory, Hawaii, U.S.A.).  Data will be released through WISEREP.}
\end{table*}

\section{Observations and data reduction} \label{sec:dataredu}
Photometry of the transient was mostly obtained using the Samuel Oschin telescope (P48) with the ZTF camera \citep[][]{2020PASP..132c8001D} in $g$ and $r$ bands.
Additional photometry was obtained with the Nordic Optical Telescope (NOT) using the Alhambra Faint Object Spectrograph and Camera (ALFOSC\footnote{\url{http://www.not.iac.es/instruments/alfosc/}}).
P48 frames were obtained through the NASA/IPAC Infrared Science Archive\footnote{\url{https://irsa.ipac.caltech.edu/Missions/ztf.html}}, while magnitudes for these data were obtained using the dedicated pipeline {\sc SNOoPY}\footnote{\url{http://graspa.oapd.inaf.it/snoopy.html}} performing point-spread-function (PSF) photometry on template subtracted images.
Templates and magnitudes of the reference stars were obtained from the Sloan Digital Sky Survey Data Release 14 \citep[DR14;][]{2018ApJS..235...42A} available through the SDSS Catalog Archive Server (CAS\footnote{\url{https://www.sdss.org/dr14/data_access/tools/}}). 
Three additional $gri$ epochs were obtained using the Liverpool Telescope \citep[LT;][]{Steele2004} with the optical imaging component of the Infrared-Optical (IO) suite of instruments (IO:O\footnote{\url{https://telescope.livjm.ac.uk/TelInst/Inst/IOO/}}).
LT data reduction was performed through a dedicated pipeline, using PSF photometry obtained on template subtracted images.
Templates and magnitudes of the photometric standards used were provided by the Pan-STARRS1 (PS1) survey \citep{2012ApJ...750...99T}.
One point of $i-$band photometry (on 2018 December 24.36~UT) was obtained using the Palomar 60-inch telescope (P60) with the SED Machine \citep[SEDM;][]{2012SPIE.8446E..86B,2018PASP..130c5003B} and reduced using the {\sc FPipe} pipeline \citep{2016A&A...593A..68F}.

A log of the spectroscopic observations is reported in Table~\ref{table:18ijpspeclog}, including the names of the instruments used and basic information about the spectra.
The classification spectrum, along with three additional spectra, were obtained with the Keck-I telescope using the Low Resolution
Imaging Spectrograph \citep[LRIS;][]{1994SPIE.2198..178O} and reduced using the automated pipeline {\sc LPipe} \citep{2019PASP..131h4503P}. 
Three additional spectroscopic observations were performed using the NOT with ALFOSC, reduced using {\sc foscgui}\footnote{\url{http://graspa.oapd.inaf.it/foscgui.html}}.
An additional intermediate resolution spectrum was obtained using the ESO Very Large Telescope (VLT) with the X-shooter \citep{2011A&A...536A.105V}, echelle spectrograph\footnote{obtained under programme 0102.D-0221 (P.I. Sollerman).}
and reduced using the ESO dedicated pipeline through the {\sc esorex v3.2.0} \citep{2015ascl.soft04003E} and {\sc gasgano} environments.

\section{Analysis and discussion} \label{sec:analysis}
\begin{figure*}
\begin{center}
\includegraphics[width=\linewidth]{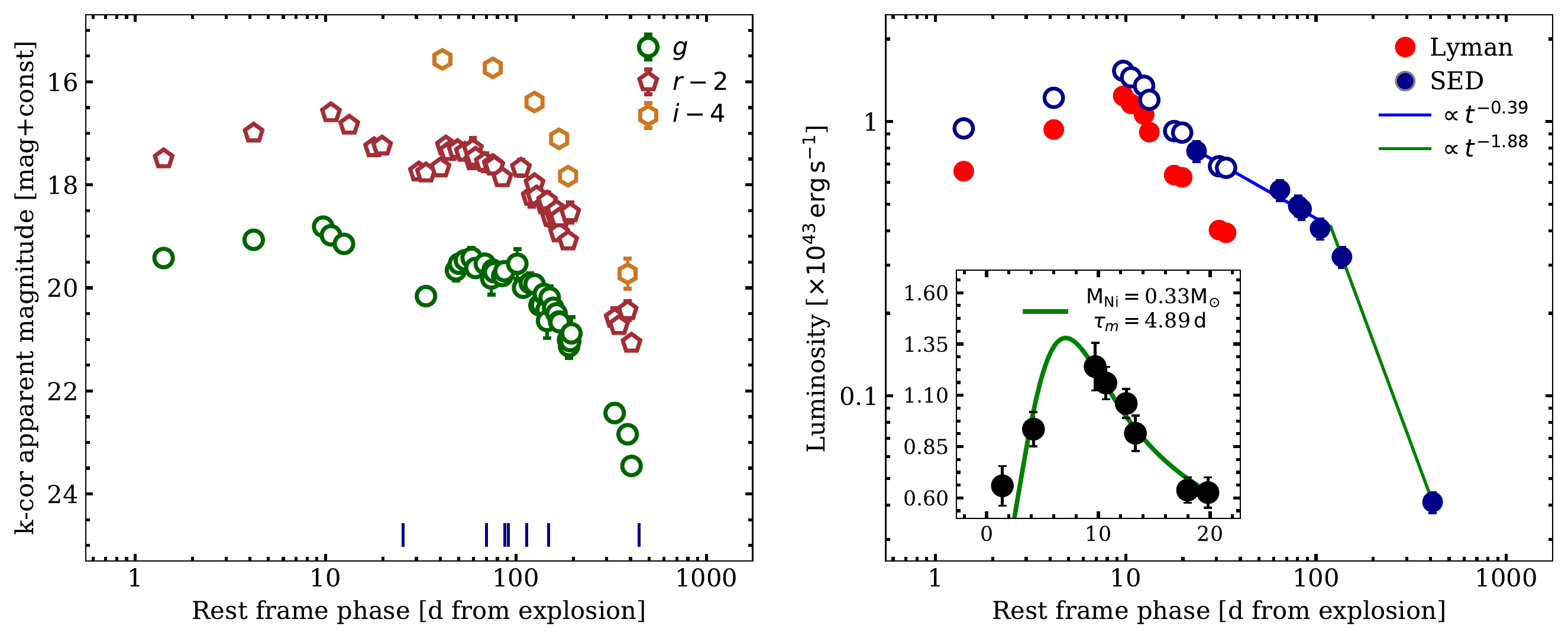}
\caption{{\bf Left:} The $gri$ light curves of \object, k-corrected apparent magnitudes versus  rest frame days. Blue ticks at the bottom mark the epochs of spectroscopic observations. {\bf Right:} Bolometric light curve of \object~estimated following the prescriptions of \citet{2014MNRAS.437.3848L} (red points) and by fitting the SEDs obtained from the spectra (solid blue points). An estimate of the early peak obtained matching the luminosities obtained at $\simeq+34\,\rm{d}$ is also shown (open blue points). The inset shows the fit of the model of \citet{1982ApJ...253..785A} to the early luminosity evolution of \object, resulting in $M_{Ni}\simeq0.3$\msun~and $\tau_m\simeq4.9\,\rm{d}$, corresponding to $E_k\simeq3.3\times10^{51}\,\rm{erg}$ and $M_{ej}\simeq0.7$\msun. \label{fig:lightcurves}}
\end{center}
\end{figure*}
\subsection{Photometry} \label{sec:photometry}
Pre-SN observations of the field of \object~were obtained by ZTF since 2018 March 31.3~UT, resulting in no detections down to average magnitudes of $\simeq21\,\rm{mag}$ in both $g$ and $r$ bands.
Last non-detection limits were obtained on 2018 November 4.5~UT (corresponding to $g>20.6$ and $r>21.4\,\rm{mag}$), roughly three days before the first $g-$ and $r-$band detections.
We will therefore adopt 2018 November 6.0~UT ($\rm{JD}=2458428.5\pm1.5$, the mid-point between the first detection and the last non-detection limit) as an estimate of the explosion epoch of \object~and refer to phases with respect to this date.
Magnitudes at rest frame epochs were obtained computing k-corrections using the spectra of \object~and following the prescriptions of \citet[][see their Eq.~13]{2002astro.ph.10394H}, adopting a recessional velocity of $cz=25540$\kms, as derived from the redshift estimated from the host lines in the X-shooter spectrum ($z=0.0852$; see Sect.~\ref{sec:xshooter}).
The resulting $gri$ light curves are shown in Fig.~\ref{fig:lightcurves} (left panel), along with an estimate of the bolometric luminosity of \object~(right panel), which will be discussed below.

The early photometric evolution is fast, with both $g-$ and $r-$band light curves rapidly rising to a first maximum within $\simeq10\,\rm{d}$ from the SN explosion.
At $\simeq+34\,\rm{d}$ both the $g-$ and $r-$band light curves show a further rise to a second and broader peak (lasting $\simeq25\,\rm{d}$), while the $i-$band light curve does not reveal the same `double-peaked' shape due to lack of early observations in this band.
After the second peak ($t\gtrsim60\,\rm{d}$), the photometric evolution is slower in all bands, with decline rates of $\simeq0.012$, $0.011$ and 
$0.013\,\rm{mag}\,\rm{d^{-1}}$ in $g-$, $r-$ and $i-$band, respectively.
The rise times in $g$ and $r$ were computed fitting the early evolution in each band with a second-order polynomial in order to estimate the epoch of the maximum in each band.
These were computed as $t_{max}-t_{expl}$, giving  $t_{rise,g}=7.9\pm1.5$ and $t_{rise,r}=10.7\pm1.5\,\rm{d}$, where the errors are completely dominated by the uncertainty on the explosion epoch.
These rise times are slightly faster than the average $r-$band rise time inferred by \citet{2015A&A...574A..60T} and \citet{2019A&A...621A..71T} for their samples of SNe Ic-BL ($\simeq14.7$ and $\simeq15\,\rm{d}$, respectively), although still comparable with the faster end of the SN Ic-BL iPTF rise-time distribution presented in \citet[][their fig. 17]{2019A&A...621A..71T}.
We note, in addition, that the photometric evolution during the rise is well reproduced by power-laws of the form $L_g\propto t^{0.29}$ and $L_r\propto t^{0.41}$\ergs, with fluxes in $g$ and $r$ computed using the zero-points for the ZTF filters reported by the Spanish Virtual Observatory \citep[SVO\footnote{\url{http://svo2.cab.inta-csic.es/svo/theory/fps3/}};][]{2012ivoa.rept.1015R}.

The $g-r$ early (i.e., at $t\lesssim 10\,\rm{d}$) color evolution is relatively fast, with the color index increasing from $\simeq0.15$ to $\simeq0.60\,\rm{mag}$.
At $t\gtrsim10\,\rm{d}$, the $g-r$ index evolves toward bluer colors until $\simeq+80\,\rm{d}$, thereafter remaining roughly constant ($\simeq0.1\,\rm{mag}$) throughout the rest of the photometric evolution of \object.
At $t\gtrsim+60\,\rm{d}$, we note an almost linear decline in $r-i$, with the color index becoming progressively bluer with time, as reflected by the evolution of the pseudo-continuum observed in the spectra of \object~(see Sect.~\ref{sec:spectroscopy}).

Absolute magnitudes were obtained, after correcting apparent values for the Galactic reddening and adopting a distance modulus $\mu=37.85\,\rm{mag}$ (see Sect.~\ref{sec:intro}).
The resulting $M_{r,peak}$ falls within the brighter end of the distribution of peak magnitudes presented in \citet{2019A&A...621A..71T}.

\subsubsection{Evolution of the bolometric luminosity} \label{sec:bolom}
The early ($t\lesssim+30\,\rm{d}$) bolometric light curve of \object~was obtained following the prescriptions of \citet[][see their Eq.~6 for their sample of SE SNe]{2014MNRAS.437.3848L}, allowing an estimate of the $g-$band bolometric corrections from the evolution of the $g-r$ color.
The resulting light curve peaks at $\simeq1.24\times10^{43}$\ergs, with the maximum occurring at $t^{bol}_{peak}\simeq+9.7\,\rm{d}$.
The total radiated energy is $\simeq2.1\times10^{49}\,\rm{erg}$ within the first $34\,\rm{d}$.

An alternative estimate of the luminosity can be obtained using the information on the spectral energy distribution (SED) available through the analysis of the spectra at $t\ge+24\,\rm{d}$.
We computed $BVRI$ and $gri$ synthetic photometry using the {\sc calcphot} task available through the IRAF/STSDAS Synthetic Photometry ({\sc synphot}) package and fitted black body (BB) functions to the resulting SEDs.
Final luminosities were then obtained integrating the fluxes in each band excluding the spectral region at wavelengths shorter than 2000~\AA, where the flux is expected to be suppressed by line blanketing \citep[see, e.g.,][]{2017ApJ...850...55N}.
Assuming a power-law decline at $t\gtrsim30\,\rm{d}$ (see Fig.~\ref{fig:lightcurves}; right panel), and interpolating the bolometric light curves at the same phases, we estimated an offset of $2.85\times10^{42}$\ergs~between the two methods, as computed at $+34\,\rm{d}$.
Applying this offset to the early light curve would give a peak luminosity of $\simeq1.5\times10^{43}$\ergs~with a total radiated energy of $\simeq2.8\times10^{49}\,\rm{erg}$ within the first $34\,\rm{d}$ and $\simeq1.4\times10^{50}\,\rm{erg}$ during the $410\,\rm{d}$ covered by our follow-up campaign.

A comparison of the $+24\,\rm{d}$ spectrum with the SN templates included in the SuperNova IDentification \citep[SNID\footnote{\raggedright\url{https://people.lam.fr/blondin.stephane/software/snid/}};][]{2007ApJ...666.1024B} tool and the GEneric cLAssification TOol \citep[GELATO\footnote{\url{https://gelato.tng.iac.es/gelato/}};][]{2008A&A...488..383H}, gives a good match with SNe Ic/Ic-BL, with a particular good match with SNe~2004aw and 1998bw around peak (see Sect.~\ref{sec:classification}).
In the following, we will therefore compare the main observables of \object~at $t\lesssim20\,\rm{d}$ with quantities inferred from samples of SNe Ic-BL, and conduct simple modeling of such an early light curve.
Following \citet[][their Eq.~4]{2016MNRAS.457..328L}, the average peak luminosity obtained with the methods described above already suggests a relatively high mass of $^{56}\rm{Ni}$ expelled by the SN explosion ($M_{\rm{Ni}}$ $\simeq0.57$\msun), although comparable to the average value found by \citet{2011ApJ...741...97D} for their sample of SNe Ic-BL.

The total mass and the kinetic energy of the ejecta can be derived following the prescriptions of \citet{1982ApJ...253..785A} (see also the formulation of \citealt{2015MNRAS.450.1295W} of the analytical model applied to a sample of SE SNe).
The model assumes spherical symmetry, a constant optical opacity $\kappa_{opt}$, small initial radius ($R_0<<10^{14}\,\rm{cm}$) and homologous expansion of the optically thick ejecta $R(t)=R_0+v_{sc}t$, with $v_{sc}$ being the scale expansion velocity \citep[see][]{1982ApJ...253..785A}. 
Under these assumptions, the characteristic time scale $\tau_m=\sqrt{2\tau_0\tau_h}$ can be defined, with $\tau_0$ and $\tau_h$ being the diffusion and hydrodynamical times, respectively \citep[see][]{2015MNRAS.450.1295W}.

The evolution of the bolometric luminosity can be expressed as a function of the kinetic energy of the ejecta $E_k$, the $^{56}\rm{Ni}$ mass $M_{\rm{^{56}Ni}}$ and the total mass of the ejecta $M_{ej}$ as follows \citep[see][]{2012ApJ...746..121C}:
\begin{eqnarray}
\label{eq:boloArnett}
L(t)=M_{\rm{^{56}Ni}}\,e^{-x^2}&\biggl[\,2\,(\epsilon_{\rm{^{56}Ni}}-\epsilon_{\rm{^{56}Co}})\mathop{\mathlarger{\int}}^x_0\xi e^{-\xi\frac{\tau_m}{\tau_{Ni}+\xi^2}}d\xi + \nonumber \\
&\epsilon_{\rm{Co}} \mathop{\mathlarger{\int}}^x_0 \xi e^{-\frac{\xi\tau_m}{\tau_{Ni}}\left(1-\frac{\tau_{Co}-\tau_{Ni}}{\tau_{Co}\tau_{Ni}}\right)+\xi^2}d\xi\biggr],
\end{eqnarray}
where $x\equiv t/\tau_m$, $\epsilon_{Co}=6.78\times10^9\,\rm{erg}\,\rm{s^{-1}}\,\rm{g^{-1}}$ and $\epsilon_{Ni}=3.90\times10^{10}\,\rm{erg}\,\rm{s^{-1}}\,\rm{g^{-1}}$ \citep[see, e.g.,][]{1997A&A...328..203C} and $\tau_{Co}$, $\tau_{Ni}$ are the radioactive decay times of $^{56}\rm{Co}$ and $^{56}\rm{Ni}$ \citep[$111.3$ and $8.8\,\rm{d}$, respectively; see, e.g.,][]{1994ApJS...92..527N}.
Assuming a constant optical opacity $\kappa_{opt}=0.07\,\rm{cm^2}\,\rm{g^{-1}}$ \citep{2000AstL...26..797C} and fitting Eq.~\ref{eq:boloArnett} to the bolometric light curve of \object~gives $M_{^{56}Ni}=0.33\pm0.05$\msun~and $\tau_m=4.89\pm0.80\,\rm{d}$.
Assuming a uniform density within the ejecta, $\tau_m$
can also be expressed as follows:
\begin{equation}
\tau_m=\left(\frac{2\kappa_{opt}}{\beta c}\right)^{1/2}\left(\frac{3\,M^3_{ej}}{10\,E_k}\right)^{1/4},
\label{eq:taum}
\end{equation}
where $\beta$ is an integration constant ($\beta\simeq13.8$, as in \citealt{2015MNRAS.450.1295W}).
The degeneracy between the kinetic energy and the total mass of the ejecta
$E_k=1/2 M_{ej}<v^2>$, with $<v^2>$ being the mean squared expansion velocity, can be broken assuming uniform density within the expanding ejecta and hence $<v^2>=3/5v^2_{ph}$ \citep{1982ApJ...253..785A}, where $v_{ph}$ is the photospheric velocity.
An estimate of $<v>$ can be obtained measuring the minima of the P-Cygni absorption profiles of \ion{Fe}{II} or \ion{O}{I} lines around maximum light \citep[see, e.g.][]{2016MNRAS.458.1618D}, which, in the case of \object, corresponds to $v_{\ion{O}{I}}\simeq12400$\kms~at $\simeq+24\,\rm{d}$ (see Sect.~\ref{sec:spectroscopy}).
Following \citet{2016MNRAS.458.1618D}, this corresponds to $<v>\simeq21200$\kms.
Taking this value for the mean expansion velocity, Eq.~\ref{eq:taum} gives $M_{ej}=0.73\pm0.05$\msun~and $E_k=(3.26\pm0.17)\times10^{51}\,\rm{erg}$ for the total mass and the kinetic energy of the ejecta.

The derived values are similar to those found by \citet{2019A&A...621A..71T} for their sample of SNe Ic-BL, with an ejected mass comparable to that found for PTF11lbm using a similar approach. However, the $^{56}\rm{Ni}$ mass is higher, corresponding to $\sim45$\% of $M_{ej}$.
While these values do not make of \object~the most extreme case (see, e.g., the values found for iPTF16asu), the relative amount of nickel is uncomfortably high and might suggest a non-negligible contribution of ejecta-CSM interaction to the total luminosity also around the first peak of \object.
\begin{figure}
\begin{center}
\includegraphics[width=\linewidth]{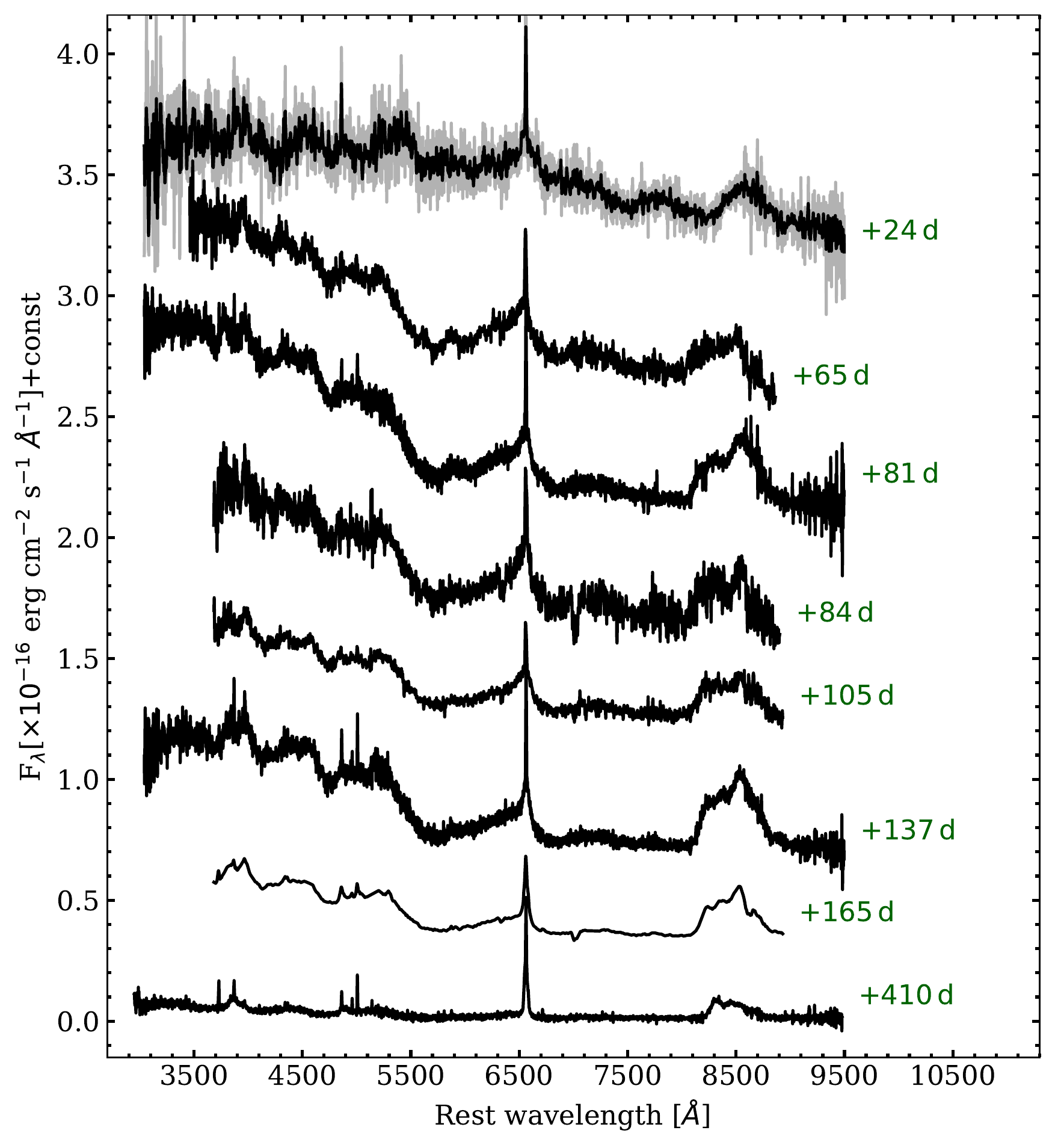}
\caption[]{Spectral sequence of \object. Rest frame phases refer to the estimated epoch of the explosion. The $+165\,\rm{d}$ spectrum has been degraded to the resolution of the $+105\,\rm{d}$ spectrum, to facilitate the comparison. The $+24\,\rm{d}$ spectrum has been degraded to 1/3 of its resolution and plotted in black, while the original spectrum is reported in gray. \label{fig:spectra}}
\end{center}
\end{figure}
\begin{figure}
\begin{center}
\includegraphics[width=\columnwidth]{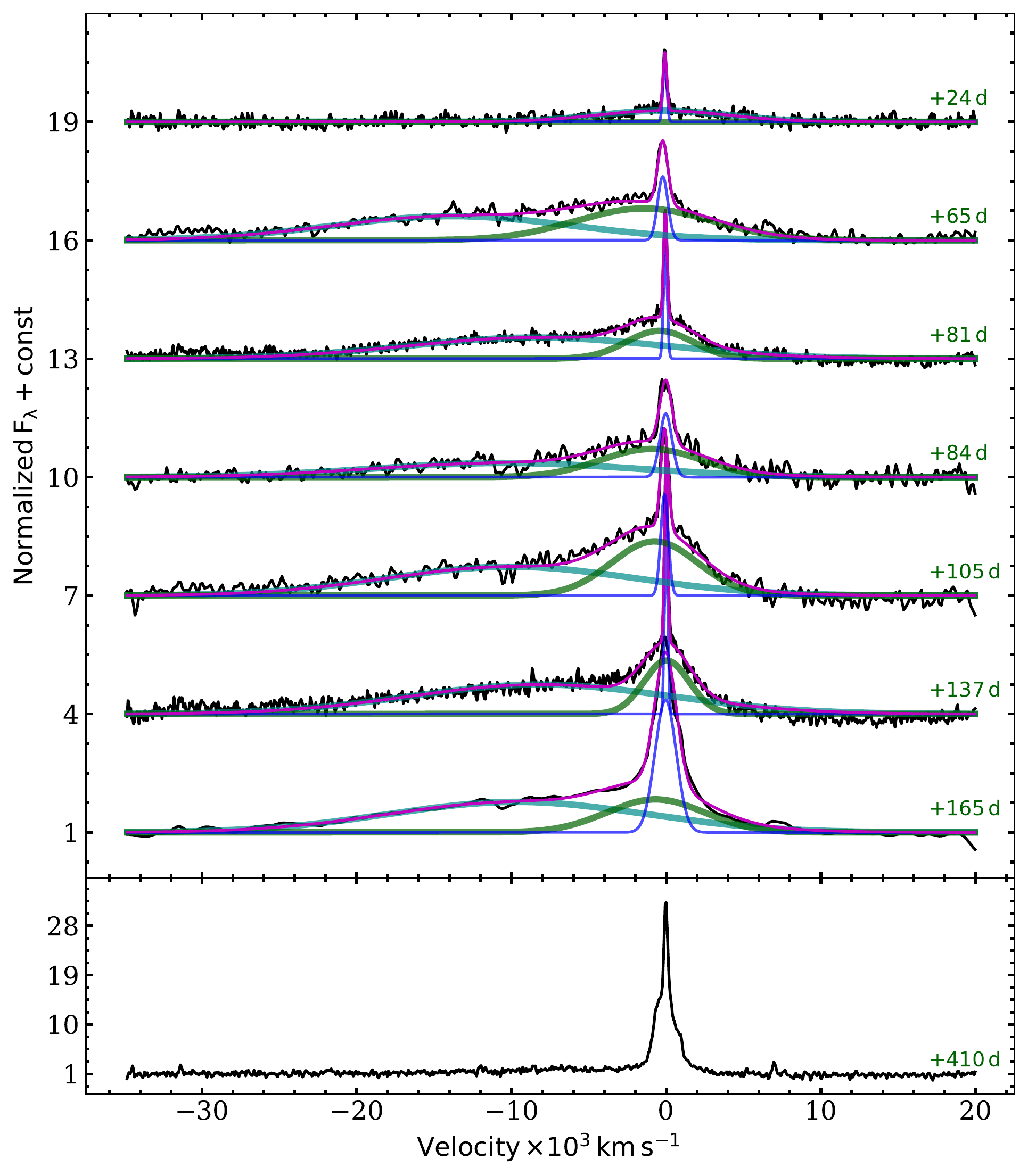}
\caption{Evolution of the \ha~profile, along with a multi-component fit. \label{fig:halpha}}
\end{center}
\end{figure}

The bolometric light curve, as constructed from the spectra, can be reproduced by a `broken power-law' starting from $+24\,\rm{d}$ (Fig.~\ref{fig:lightcurves}, right panel), with the break occurring at $\simeq+120\,\rm{d}$.
A similar behavior is observed in interacting transients, where the SN shock is expected to break through a dense and extended pre-existing CSM \citep[see, e.g.,][and references therein]{2014ApJ...797..118F,2014ApJ...781...42O,2020A&A...635A..39T}.
The total radiated energy up to $+410\,\rm{d}$, as well as the prominent narrow \ha~line visible at all phases also support a CSM interaction interpretation for the second peak of \object.
In the case of \object, at least a fraction of the luminosity output during the first peak could also be powered by interaction of the SN ejecta.
A more complex modeling, beyond the scope of this paper, is probably required to properly model the evolution of \object~in order to estimate its explosion parameters.

\subsection{Spectroscopy} \label{sec:spectroscopy}
\subsubsection{Low resolution spectroscopy} \label{sec:lowres}
Low- and medium-resolution optical spectra, calibrated against photometry obtained at the closest epochs, are shown in Fig.~\ref{fig:spectra}.
At $+24\,\rm{d}$, the spectrum shows un-resolved Balmer lines in emission (\ha~and \hb) on top of shallower broader features.
The lack of other narrow emission lines typically associated with \ion{H}{II} regions (e.g., \ion{[O}{III]}, \ion{[O}{II]} and \ion{[N}{II]}) would suggest that these are recombination lines arising from an un-shocked CSM, although the signal--to--noise (S/N) ratio of the spectrum is not sufficient to rule out the presence of such lines (see Sect.~\ref{sec:xshooter}).
Blends of \ion{Fe}{II} lines are likely responsible for the ``bumps" observed between 4000 and 5000\ang~(multiplets 26, 27, 28, 37 and 38) and at $\sim5300$\ang~(multiplets 42, 48 and 49), making a direct estimate of the ejecta photospheric velocity \citep[through the 
\ion{Fe}{II} $\lambda5169$ line;][]{2005A&A...439..671D} difficult.
On the other hand, at $\lambda\gtrsim7000$\ang~the spectrum shows broader features of \ion{O}{I} $7772-7775$\ang.
From the minimum of the \ion{O}{I} P-Cygni absorption we inferred an expansion velocity of $\simeq12400$\kms, with the blue wing extending to $\simeq2\times10^4$\kms, which, following the discussion in \citet{2016MNRAS.458.1618D}, corresponds to a mean expansion velocity of $\simeq21200$\kms.
At the same epoch, and throughout the spectroscopic evolution of \object~covered by our follow-up campaign, we also detect the near infrared (NIR) \ion{Ca}{II} triplet (see Fig.~\ref{fig:spectra}).
The $+65\,\rm{d}$ spectrum reveals a significant evolution, with the continuum becoming apparently much bluer.
The total luminosity of \ha~(measured in the $6000-7000$\ang~range) at $+65\,\rm{d}$ ($\simeq1.1\times10^{41}$\ergs) also shows a drastic increase with respect to the previous epoch ($L_{H\alpha,+24\,\rm{d}}\simeq2.8\times10^{40}$\ergs), subsequently remaining roughly constant up to $+137\,\rm{d}$.
\begin{figure*}
\begin{center}
\includegraphics[width=0.85\linewidth]{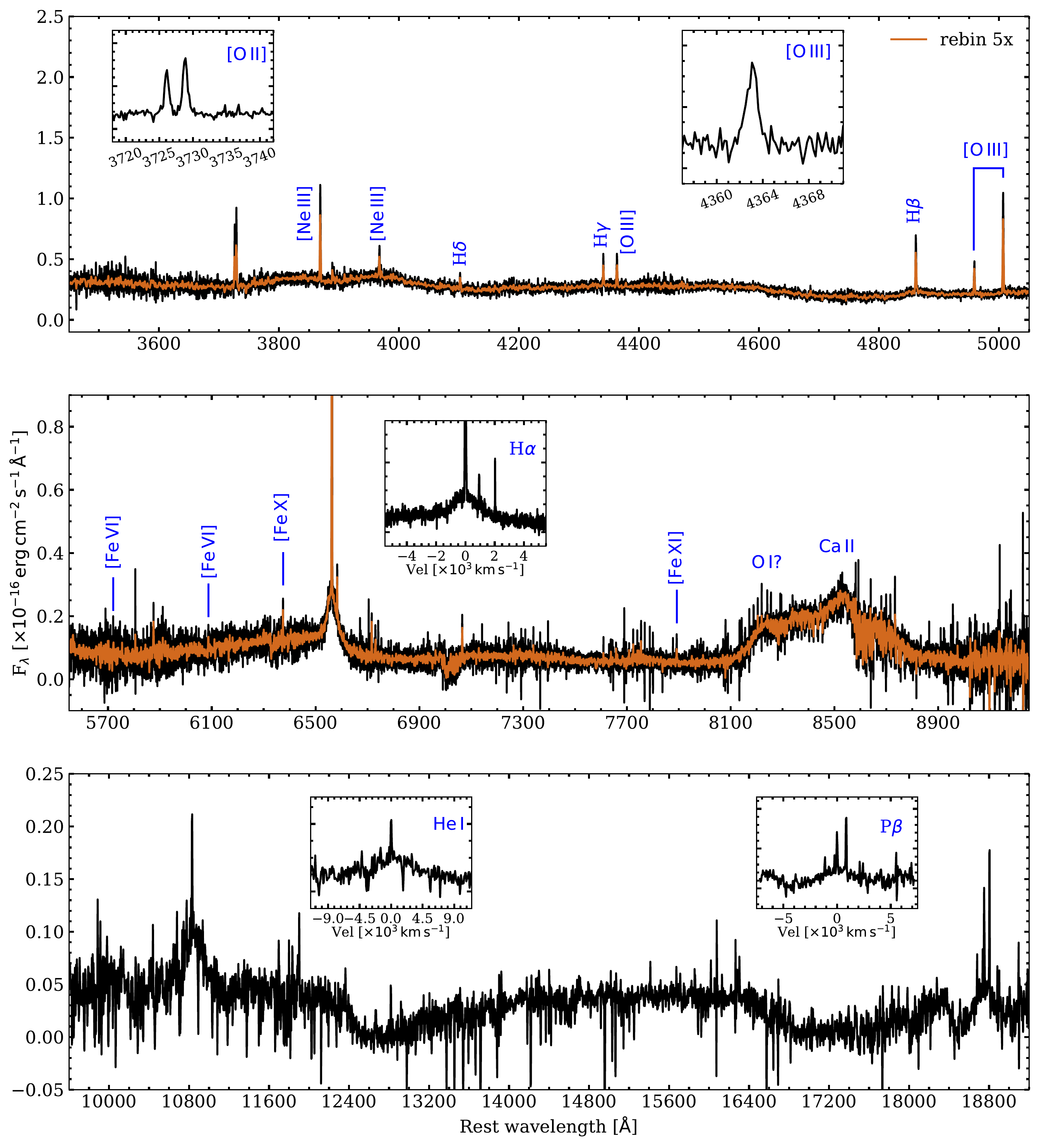}
\caption{X-shooter spectra of \object~obtained at $+165\,\rm{d}$. {\bf Top} panel shows insets with zoom-ins of the \ion{[O}{II]} $\lambda\lambda3726, 3729$  and \ion{[O}{III]} $\lambda4363$ regions. {\bf Middle} and {\bf bottom} panels show insets of the \ha, \ion{He}{I} $\lambda10830$ and Pa$\alpha$ regions in velocity space. The NIR spectrum has been re-rebinned to a fifth of its resolution to facilitate the identification of the main emission features. \label{fig:xshooter}}
\end{center}
\end{figure*}
While blue excesses can be generally associated with the contribution of fluorescence from numerous blended Fe lines \citep[see, e.g.,][and references therein]{2020A&A...635A..39T}, an increase in the temperature of the pseudo-continuum can also be interpreted as a result of ongoing ejecta-CSM interaction.
This interpretation would also be supported by the shape of the bolometric light curve (Fig.~\ref{fig:lightcurves}, right panel), showing a ``broken power-law" shape typical of interacting SNe.
A ``delayed" interaction might be explained by the presence of a confined dense shell surrounding the progenitor star of \object.
Based on the evolution of the bolometric luminosity and the results discussed in Sect.~\ref{sec:bolom}, as well as on the \ha~evolution discussed below, we can assume that the onset of ejecta-CSM interaction is at $t\gtrsim+24\,\rm{d}$.
A constant expansion velocity of $v\simeq21200$\kms~(see above) would place the shell at a distance of $\simeq4.4\times10^{15}\,\rm{cm}$ from the progenitor star of \object.
This estimate is of the same order of magnitude as that inferred from the BB fit performed at $+65\,\rm{d}$ ($\simeq10^{15}\,\rm{cm}$).
Assuming that the SN shock breaks through the confined shell roughly at $t\simeq120\,\rm{d}$ (i.e., when the break in the power-law describing the light curve occurs), this would imply an external radius of the shell of $\simeq2.2\times10^{16}\,\rm{cm}$.
Such a detached H-rich shell might be produced by a single massive progenitor and hence expelled $20-100\,\rm{yr}$ before CC, assuming a$v_{wind}\simeq70$\kms, as estimated in Sect.~\ref{sec:xshooter} \citep[see also][for a similar interpretation for SN~2014C]{2017ApJ...835..140M}, or the last eruptive episode of a Wolf-Rayet star in its transitional phase from the LBV phase \citep[see, e.g., the case of the Type Ibn SN~2006jc;][]{2007Natur.447..829P}, but could also originate from a H-rich companion through a stationary wind or an eruptive event (see Sect.~\ref{sec:conclusions}).

In Fig.~\ref{fig:halpha} we show the evolution of the spectral region around \ha, revealing a structured and asymmetric profile throughout the spectroscopic evolution of the transient.
At $+24\,\rm{d}$, the overall profile is well reproduced by a broad, blue-shifted component with a $\rm{FWHM}\simeq10^4$\kms~and a narrow unresolved component ($\rm{FWHM}<70$\kms) centered at \ha~rest wavelength, although a comparison with early spectra of the Type Ic-BL SN~1998bw might suggest a different interpretation for this feature (see Sect.~\ref{sec:classification}).
A multi-gaussian fit between $+65$ and $+165\,\rm{d}$ reveals two broad components: a blue-shifted component with a FWHM of $\simeq2\times10^4$\kms~and a redder one with a FWHM slowly decreasing from $\simeq10^4$ to $\simeq3\times10^3$\kms, with a third narrow and unresolved component.
While the redder component is likely due to the wings of the typical electron scattering profile observed in high-resolution spectra of interacting transients \citep[see, e.g.,][]{2018MNRAS.475.1261H}, the corresponding velocity width is also consistent with those observed in shocked regions of dense media typically surrounding the progenitors of SNe IIn.
A clumpy \citep[e.g.,][]{1994MNRAS.268..173C} or highly asymmetric \citep[e.g.,][]{2014MNRAS.438.1191S} CSM, could explain the simultaneous presence of a broad component, which would then be produced by the outer ionized layers of the freely expanding SN ejecta and an intermediate component arising from the shocked CSM, with the narrow emission feature possibly produced in the ionized un-shocked CSM \citep[see, e.g.][although see Sect.~\ref{sec:xshooter} for a possible different interpretation of the \ha~profile observed at $t\ge+65\,\rm{d}$]{1993MNRAS.262..128T}.
By $+410\,\rm{d}$, the shape of \ha~has changed significantly, showing a narrow component on top of a boxy, flat-topped line profile, reminiscent of the overall profile observed in H-rich interacting SNe IIn \citep[see, e.g.,][and the discussion in Sect.~\ref{sec:conclusions}]{2020A&A...638A..92T} and is not well reproduced by a combination of Gaussian profiles.

\subsubsection{The X-shooter spectrum} \label{sec:xshooter}
Medium resolution spectra were obtained with X-shooter on 2019 May 4.06~UT ($\rm{JD=2458607.56}$, $t=+165\,\rm{d}$).
The observations consist of 4 exposures per arm, median combined to produce a single spectrum covering the $350-2000\,\rm{nm}$ wavelength range\footnote{\raggedright\url{https://www.eso.org/sci/facilities/paranal/instruments/xshooter/overview.html}}.
Each observation was obtained at airmass $\lesssim1.5$, with an average seeing of $0\farcs7$ and we therefore take the nominal values of resolution for each arm ($\rm{R}\equiv\lambda/\Delta\lambda=5400$, 8900 and 5600 in the UVB, VIS and NIR arm, respectively, for slit widths of $1\farcs0$ in UVB and $0\farcs9$ in VIS and NIR)\footnote{\raggedright\url{https://www.eso.org/sci/facilities/paranal/instruments/xshooter/inst.html}}.
These are consistent with the resolutions measured from the available sky lines.
The resulting spectra are shown in Fig.~\ref{fig:xshooter}.

The \ha~region shows a structured profile with a narrow ($\rm{FWHM}\simeq70$\kms) and an ``intermediate" ($\rm{FWHM}\simeq7400$\kms) component, both centered at \ha~rest wavelength, on top of a broader ($\rm{FWHM}\simeq21500$\kms) blue-shifted ($v_{shift}\simeq10600$\kms) component (see also Fig.~\ref{fig:halpha}).
A much shallower broad component is observed in \hb, with a FWHM of $\simeq2130$\kms~and an unresolved narrow component ($\rm{FWHM}<60$\kms.)

The measured \hd/\hb~and \hg/\hb~line ratios of the narrow components are $\simeq0.2$ and 0.5, roughly corresponding to the predicted Balmer decrement in a Case B recombination scenario \citep[assuming $T=10^4\,\rm{K}$ and $n_e=10^2\,\rm{cm^{-3}}$; see][]{2006agna.book.....O}, consistent with negligible contribution of the local environment to the total extinction in the direction of \object.
The electron density was inferred from the \ion{[O}{II]} line ratio \citep[$j_{3729}/j_{3726}=1.26$, corresponding to $n_e\simeq10^2\,\rm{cm^{-3}}$, see][]{2006agna.book.....O}, as well as from the \ion{[S}{II]} $j_{6716}/j_{6732}$ ratio, giving the same result.
The narrow \ha/\hb~ratio of $\simeq3.5$ is somewhat higher than that predicted in the case B scenario, which might suggests some contribution from the transient to \ha.
A number of other narrow lines are also detected, including \ion{[O}{II]} $\lambda\lambda3726,\,3729$, \ion{[O}{III]} $\lambda\lambda4363$, 4958 and 5007, \ion{[Ne}{III]} $\lambda\lambda3868$, 3967, \ion{[S}{II]} $\lambda\lambda6716$, 6732 and \ion{[N}{II]} $\lambda\lambda6548,\,6583$, which, along with the narrow Balmer lines are all below the resolution limit, with the exception of \ha~narrow component, showing a $\rm{FWHM}\simeq70$\kms.

Following the line identification of \citet{2002ApJ...572..350F}, we also detect unresolved high ionization narrow lines, such as \ion{[Fe}{XI]} $\lambda7892$, \ion{[Fe}{X]} $\lambda6375$, \ion{[Fe}{VI]} $\lambda$5720, 6087 and 5276 as well as narrow \ion{He}{I} $\lambda3889$ and 7065.

As highlighted in Figs.~\ref{fig:spectra} and \ref{fig:xshooter}, the non-detection of such features at earlier phases is most likely due to the lower S/N and resolution of the spectra.
As in SN~1995N \citep[see also][]{2020A&A...643A..79S}, these features can be interpreted as arising from the dense circumstellar gas.
These lines are significantly fainter at $+410\,\rm{d}$ where the S/N and resolution of the spectrum is still sufficient to detect such features.

The \ion{[O}{III]} $\lambda4363$ is clearly detected in the X-shooter spectrum. 
This line is usually faint compared to \ion{[O}{III]} $\lambda\lambda4959$, 5007, with ($j_{5007}+j_{4959})/j_{4363}\gtrsim50$ in typical \ion{H}{II} regions. 
The inferred value ($j_{5007}+j_{4959})/j_{4363}=7.2$ implies very high temperatures and densities for the emitting gas \citep[e.g., $T_e\gtrsim2.7\times10^4\,\rm{K}$ for electron densities $n_e\gtrsim10^6\,\rm{cm^{-3}}$ in a 5-level atom approximation; see][]{1987JRASC..81..195D,1995PASP..107..896S}, suggesting a circumstellar origin for the \ion{[O}{III]} lines as well.
This could imply a \ion{[O}{III]} flux arising from different regions (i.e., both from an underlying \ion{H}{II} region and the CSM).
Alternatively, narrow features, such as \ion{[O}{II]}, \ion{[O}{III]} and \ion{[S}{II]}, might all have circumstellar origin, along with Fe high ionization features, with different lines mapping regions with different densities.
A similar behavior is observed in \ion{H}{II} regions, where \ion{[O}{II]} lines are thought to arise from outer regions of the nebula, whereas \ion{[O}{III]} features are produced from more uniformly distributed gas \citep[see, e.g.,][and references therein]{2014ApJ...790...75N}.

In order to further explore these different scenarios, we inspected the 2D X-shooter frames, which suggested a common origin for at least the \ion{[O}{II]} and the narrow Balmer lines, with spatially extended emission suggesting contamination from the environment of \object.
The \ion{[O}{III]} and the high ionization Fe lines, on the other hand, do not show a similar spatially extended emission, favoring a circumstellar origin.
The evolution of the narrow lines over the remaining $\sim300\,\rm{d}$ of our spectroscopic coverage also seems to support a circumstellar origin at least for the high ionization Fe lines, with have fluxes which decrease significantly from $+165$ to $+410\,\rm{d}$.

\section{On the nature of SN~2018ijp} \label{sec:classification}
\begin{figure}
\begin{center}
\includegraphics[width=\columnwidth]{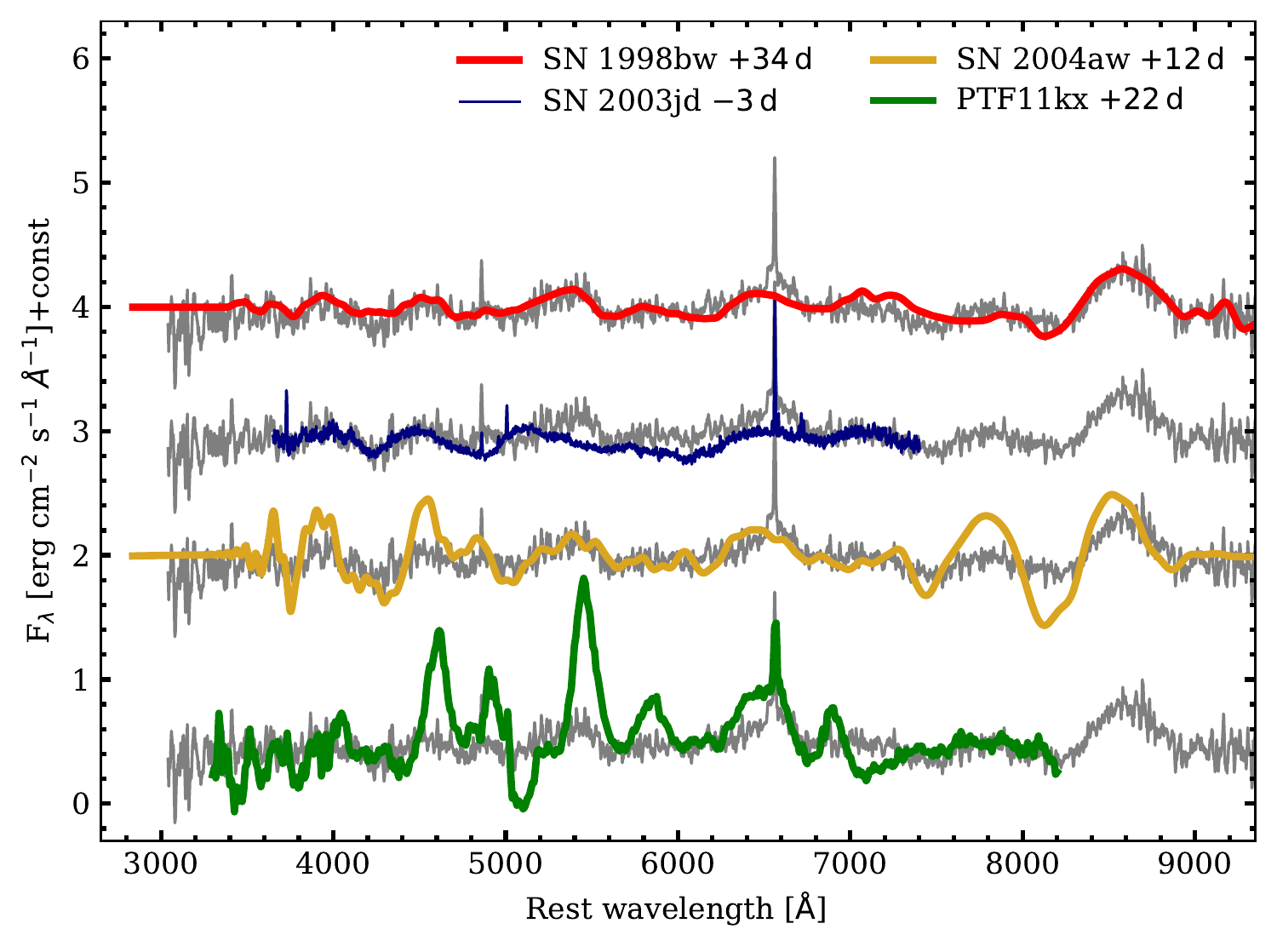}
\caption{The $+24\,\rm{d}$ spectrum ($16\,\rm{d}$ after $r-$band maximum, shown in grey) compared to ``best-match" templates obtained with the SNID (Type Ic SN~2004aw and Ic-BL SN~1998bw) and GELATO (Ic SN~2003jd) classification tools. Spectra of SNe~2004aw and 1998bw are the ``flattened" templates included in the SNID database. Phases relative to maximum light of each object are reported in the legend. The spectral continua of SNe~2018ijp and 2003jd (not included in the SNID template database) were estimated through {\it spline} functions and subtracted. The spectrum of \object~has been smoothed in order to highlight its broad features. \label{fig:specClass}}
\end{center}
\end{figure}

\begin{figure}
\begin{center}
\includegraphics[width=0.98\columnwidth]{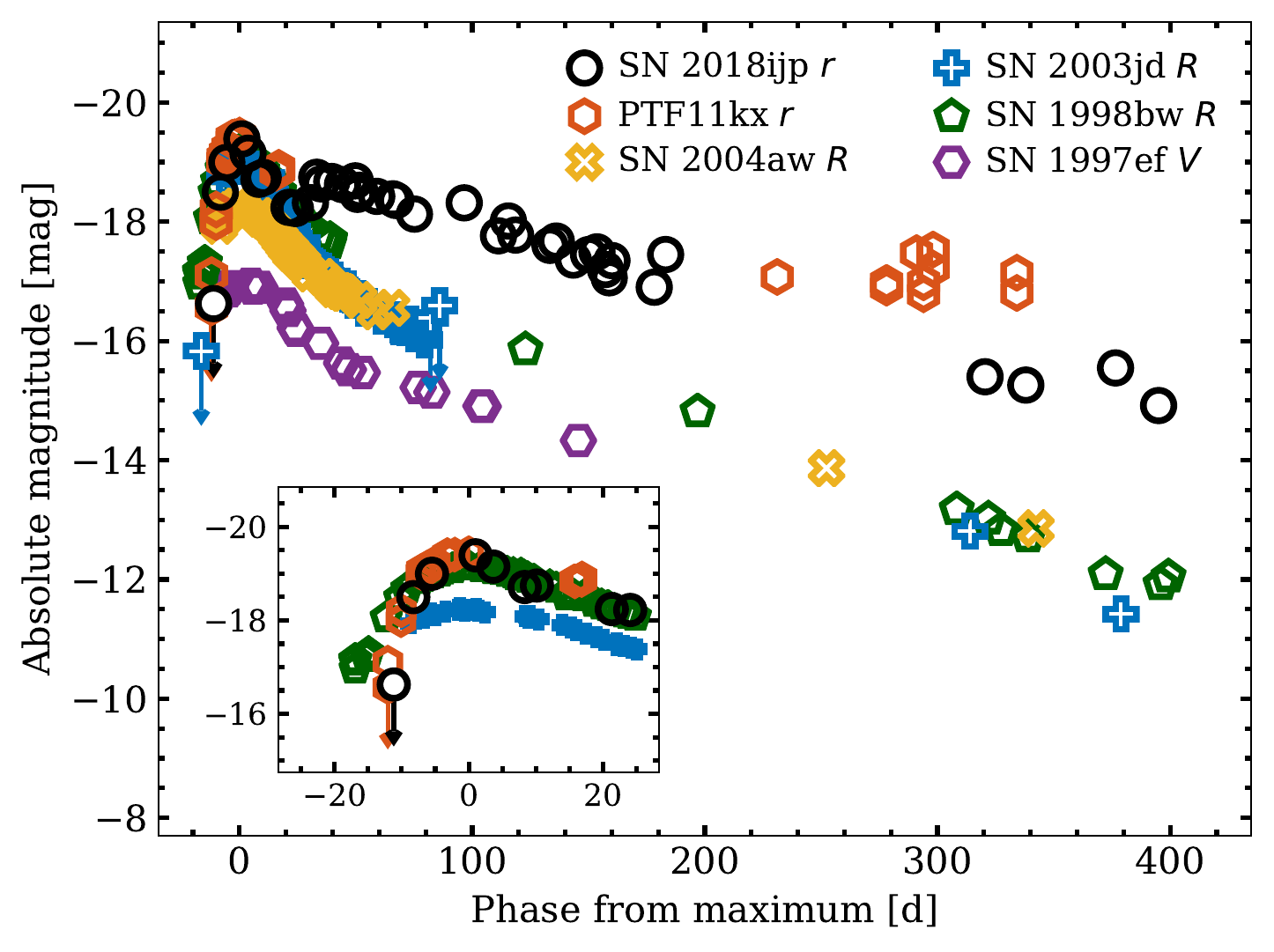}
\caption{Comparison of the absolute $r-$band light curve of \object~with those of a number of other SNe. The sample was chosen based on comparison of the $+24\,\rm{d}$ spectrum 
(see Fig.~\ref{fig:specClass}).
The choice of the reported bands is given in the legend and was made based on the available photometric data for each object. 
 In the inset, a comparison of the early $r-$band light curve of \object~with those of SNe~2003jd, 2004aw, 1998bw and PTF11kx.
\label{fig:lcComparison}}
\end{center}
\end{figure}

Since spectra at $t\ge+65\,\rm{d}$ are dominated by interaction, we use the one obtained at $+24\,\rm{d}$ (i.e. $16\,\rm{d}$ after the first $r-$band peak) to investigate the nature of the explosion of \object.
Our approach is based on automatic comparisons of the observed spectrum with the databases of SNID (based on the correlation techniques of \citealt{1979AJ.....84.1511T}) and GELATO tools.
SNID, in particular, is able to clip narrow emission lines at a specified redshift, in order to avoid spurious correlation peaks. 
Using this option in order to avoid contamination from the narrow \ha~component, we get the best match with the Type Ic SN~2004aw \citep{2006MNRAS.371.1459T} $12\,\rm{d}$ after maximum, although the comparison also suggests many other Type Ic/Ic-BL SNe as good matches to the spectral features of \object, including the Type Ic-BL SN~1998bw (see Fig.~\ref{fig:specClass}, where we also include a comparison with the Type Ia-CSM PTF11kx around peak, disfavoring such a classification for \object).
A similar result was obtained using GELATO, giving the best match with the Type Ic-BL SN~2003jd \citep{2008MNRAS.383.1485V} around maximum light.
These comparisons thus suggest a SE SN classification for \object.
In Fig.~\ref{fig:lcComparison} (also including the peculiar Ic SNe~1997ef \citealt{2000ApJ...534..660I,2000ApJ...545..407M}), on the other hand, we show that the first peak of \object~resemble the photometric evolution of PTF11kx (see the inset), with similar absolute magnitudes and fast rise to maximum, although SN~1998bw also gives a good match at $t\gtrsim-10\,\rm{d}$ (with respect to the $r-$band peak).
Around the second peak, \object~shows a different evolution with respect to all other comparison objects with the possible exception of PTF11kx, although its sampling is not as good as the one provided by our follow-up campaign between $20\,\rm{d}\lesssim t\lesssim220\,\rm{d}$.
Based on this similarity, and since Ia-CSM SNe can also produce spectra similar to \object~around the first peak, we further investigate the nature of \object~by extending the comparison to transients of different nature, as described below.

\begin{figure}
\begin{center}
\includegraphics[width=\columnwidth]{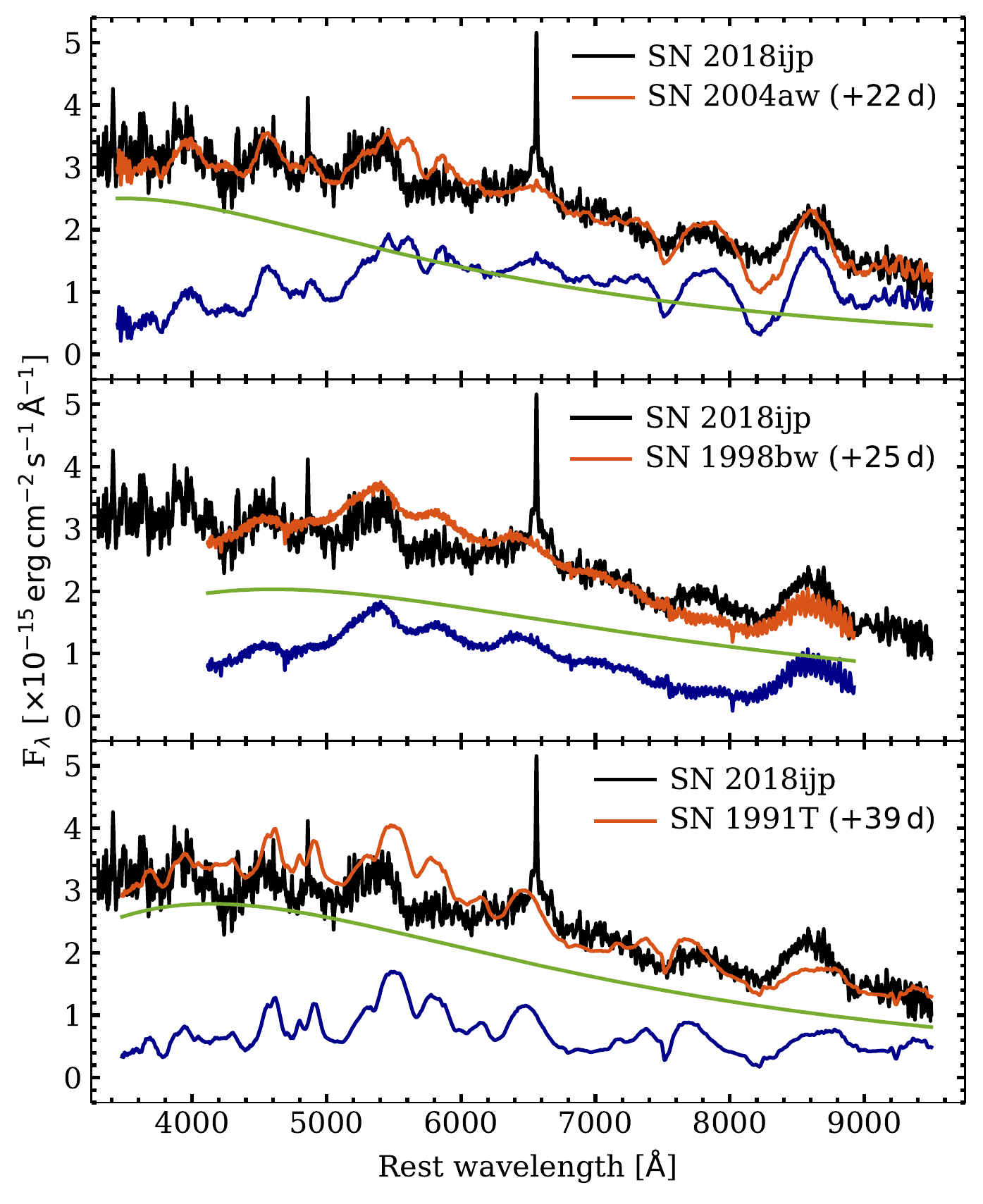}
\caption{Best fit models obtained using spectra of the Type Ic SNe~2004aw, Ic-BL 1998bw and Ia 1991T, adding a BB to the comparison objects. Blue scaled spectra are those of the comparison objects, while green lines are the BBs required to get the ``best-fit" diluted spectra (see the main text for details). The spectrum of \object~is obtained at $+24\,\rm{d}$ ($\simeq16\,\rm{d}$ after maximum, in the rest frame). Phases refer to maximum light. \label{fig:diluted}}
\end{center}
\end{figure}

Following \citet{2006ApJ...650..510A}, we compared the $+24\,\rm{d}$ spectrum of \object~(i.e., $\simeq16\,\rm{d}$ after maximum) with ``diluted" spectra of SN~1991T \citep{1992ApJ...384L..15F,1995A&A...297..509M,1998AJ....115.1096G,2012MNRAS.425.1789S} around peak.
This choice is based on the claim that SNe Ia-CSM seem to show an association with 91T-like transients \citep[see, e.g.,][]{2003Natur.424..651H,2006ApJ...650..510A,2007arXiv0706.4088P,2012A&A...545L...7T}.
Spectra of SN~1991T around maximum were diluted using a black-body in order to simulate the effects of interaction -- including a constant factor to account for the different distances and luminosities of the two objects -- and the ``best-fit model" was determined through a simple $\chi^2$ minimization routine.
A similar approach was adopted by \citet{2015A&A...574A..61L} to test how interaction would affect the appearance of SN spectra and their classification, although we did not include narrow Balmer lines (i.e., \ha) in our fit.
In Fig.~\ref{fig:diluted}, we show the results of this fitting procedure, giving a good match with SNe~2004aw and 1998bw and thus confirming the results obtained with SNID, while the best-fit spectrum obtained with SN~1991T does not seem to reproduce well the broad features observed in \object.
We note that a good match with SN~2004aw is obtained even without including a BB in the fitting procedure, suggesting that the interaction does not significantly shape the pseudo-continuum of \object~around the first peak.
This is also the rationale for our light curve analysis in Sect.~\ref{sec:bolom}.

\begin{figure}
\begin{center}
\includegraphics[width=0.96\columnwidth]{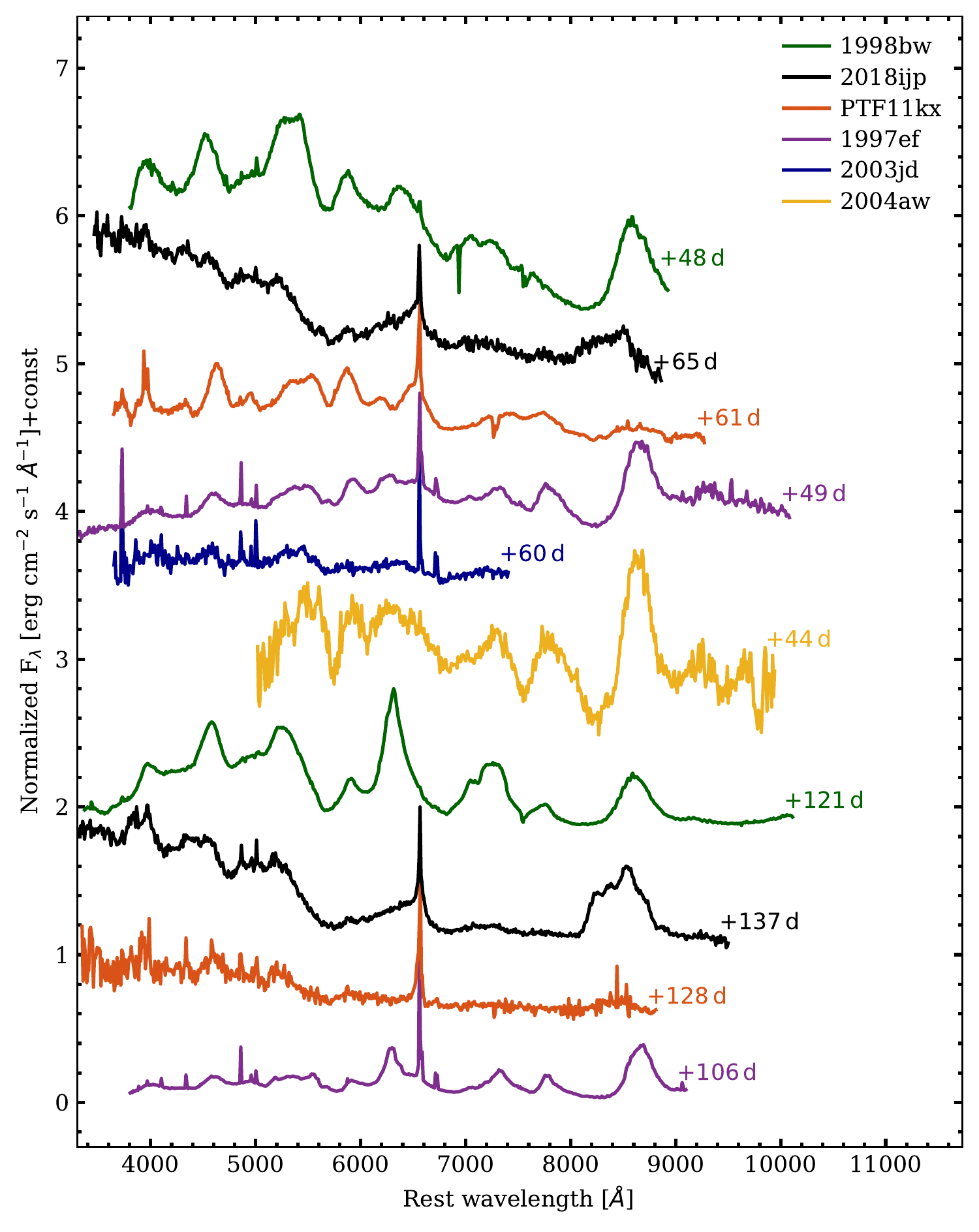}
\caption{Comparison of spectra of \object~at selected epochs and spectra of the SNe discussed in the main text at similar phases. Phases refer to the epoch of maximum light.
\label{fig:specComparison}}
\end{center}
\end{figure}

\begin{figure}
\begin{center}
\includegraphics[width=\columnwidth]{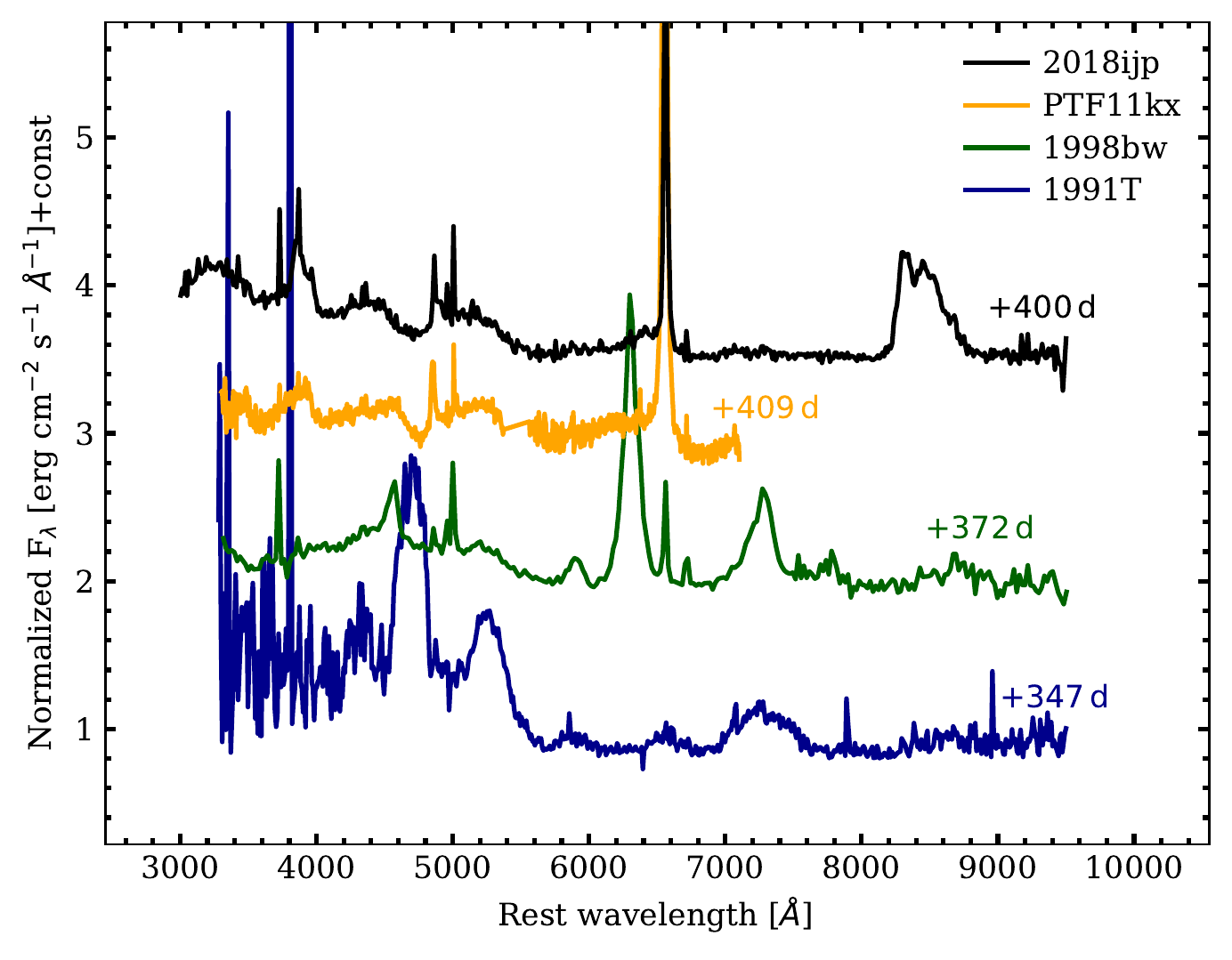}
\caption{Comparison of the late spectrum of \object~obtained at $+410\,\rm{d}$ ($\simeq400\,\rm{d}$ after $r-$band maximum) with spectra of the SNe discussed in the main text. Phases refer to the epochs of maximum light. \label{fig:nebular}}
\end{center}
\end{figure}

At later phases (i.e. $t\ge+65\,\rm{d}$), we could not find a good match with any template spectrum either using SNID or GELATO, although, after clipping the narrow \ha~component, GELATO suggests a marginal similarity with the Type Ib SN~2009er.
The same result was obtained also diluting spectra of SNe~1991T, 1998bw and 2004aw, as detailed above.
At the same time, we also note that the late spectral evolution of \object~is different than that of the aforementioned transients (see Fig.~\ref{fig:specComparison}), although, at $+410\,\rm{d}$, we note a remarkable similarity with a late spectrum of PTF11k obtained at a similar phase (see Fig.~\ref{fig:nebular}).
These evidences support our claim that later spectra are dominated by strong signatures of interaction (namely a very blue continuum, and the presence of several \ion{Fe}{II} multiplets, with a prominent and structured \ha~line in emission; see Fig.~\ref{fig:halpha}).

\section{Summary and conclusions} \label{sec:conclusions}
We presented the peculiar photometric and spectroscopic evolution of \object~and discussed its observables.
The transient shows double-peaked $g-$ and $r-$band light curves with a fast-evolving first peak showing an evolution similar to those observed in the Type Ic-BL SN~1998bw and the Ia-CSM PTF11kx.

A comparison of the first spectrum, obtained at $+24\,\rm{d}$ (around the first peak), with spectral templates included in commonly used classification tools (such as SNID and GELATO) 
favor a SE SN, providing good matches with the peculiar Type Ic SN~2004aw and the Ic-BL SN~1998bw (see Fig.~\ref{fig:spectra}).
We could not find a good match with any SN Ia either with SNID or GELATO, or modeling the effects of ongoing ejecta-CSM interaction with a BB (see Sect.~\ref{sec:classification} and Fig.~\ref{fig:diluted}), which did not give a good match with a 91T-like SN, generally associated to SNe Ia-CSM \citep[e.g.,][]{2015A&A...574A..61L}.
We note that the ``template" spectrum of SN~2004aw $\simeq22\,\rm{d}$ after maximum resemble that of \object~even without introducing an extra component.
Based on this early spectral similarity, we model the early evolution of the bolometric luminosity of \object~using the simple analytical model of \citet{1982ApJ...253..785A}, suggesting a mass of expelled radioactive $^{56}\rm{Ni}$ of $\simeq0.3$\msun~with respect to a total ejected mass of $0.7$\msun~and kinetic energy of $3.3\times10^{51} \rm{erg}$.
This makes a reasonable match to SE SNe, with inferred values overall consistent with those found by \citet{2019A&A...621A..71T} for their sample of SNe Ic-BL, and the mean nickel mass of the iPTF Ic SN sample \citep[0.2\msun;][]{2020arXiv201008392B}.
We remark that this analysis is only meant to infer rough estimates and the resulting values have to be considered approximations of the real quantities.
In addition, the very high fraction of radioactive $^{56}\rm{Ni}$ to total ejected mass inferred for \object~might suggest that interaction also plays a role in its early evolution, as possibly indicated by the presence of narrow emission lines superimposed on Ic-BL spectral features.
In order to explain simultaneous signatures of interaction and SN features observed in the $+24\,\rm{d}$ spectrum, we need to invoke a peculiar geometrical configuration for the CSM (e.g., a highly asymmetric explosion), a ``clumpy" CSM \citep[see, e.g.,][]{1994MNRAS.268..173C}, a moderate amount of CSM (with respect to those expected in SNe IIn and Ibn) or a combination of the three.
This would, however, affect the results of our modeling of the first peak of \object, although it still gives reasonable results compared to those obtained by \citet{2019A&A...621A..71T} for their sample of SNe Ic-BL.

Inspection of the spectral region around the \ha~line (Fig.~\ref{fig:halpha}), suggests the presence of a very broad, blue-shifted component, marginally visible since $+24\,\rm{d}$ (see Fig.~\ref{fig:spectra}).
Identifying this broad feature as H would be at odds with a SE SN origin, although the comparisons with SNe Ic-BL discussed in Sect.~\ref{sec:classification} (see, e.g., Fig.~\ref{fig:specClass}) reveal that broad features at these wavelengths are also observed in SE SNe.
At $t>+24\,\rm{d}$, the spectral evolution of \object~is clearly dominated by signatures of interaction, as also highlighted by its late photometric evolution.
In this context, the very high velocities inferred for the broad \ha~component (see Fig.~\ref{fig:halpha}) at these phases could be explained with a high optical depth by incoherent electron scattering in the post-shock region.
A combined effect of a high optical depth and shock velocity was discussed by \citet[][see their fig.~22]{2020A&A...638A..92T} in order to explain the similarly structured \ha~profile of SN~2013L. 
In this scenario, while the slope of the blue wing of \ha~is strongly affected by the optical depth of the CSM, the suppression of the flux at redder wavelengths is caused by the efficient thermalization (obscuration) of the \ha~photons in the SN ejecta.
This scenario is also supported by the shape of \ha~at $+410\,\rm{d}$, showing a boxy, flat-topped profile similar to that of SN~2013L, although with lower velocities, suggesting a lower expansion velocity of the shocked material.
The match with the absolute light curve of PTF11kx at later phases, the dramatic evolution in the observed spectral continuum from $+24$ and $+65\,\rm{d}$ (see Sect.~\ref{sec:spectroscopy} and Fig.~\ref{fig:spectra}) and the shape of the bolometric light curve at $t\gtrsim25\,\rm{d}$ (see Fig.~\ref{fig:lightcurves}, right panel) seem all to support a scenario where the late-time evolution of \object~is dominated by ejecta-CSM interaction.

We therefore interpret the overall evolution of \object~invoking two main phases, a first one dominated by radioactive decays, and an interaction-dominated phase where the SN ejecta collide with a H-rich pre-existing CSM (see Sect.~\ref{sec:bolom}), although we cannot exclude a contribution of CSM interaction also during the first phase.
In this scenario, the early spectral features observed in \object, with high expansion velocities measured in the $+24\,\rm{d}$ spectrum, as well as the relatively high mass of radioactive $^{56}\rm{Ni}$ seem to suggest a stripped massive star, with the CSM either produced during a previous evolutionary stage or by a H-rich companion.

The location of the H-rich shell with respect to the center of the explosion and its estimated expansion velocity, on the other hand, is difficult reconcile with an outburst of a massive H-rich star subsequently exploding as a SE SN.
A constant expansion velocity of 21200\kms~for the SN ejecta would give an inner radius of $\simeq4\times10^{15}\,\rm{cm}$, with an outer layer extending up to $\simeq2\times10^{16}$ (assuming $v_{wind}\simeq 70$\kms, as inferred in Sect.~\ref{sec:xshooter}).
This would suggest that the mass-loss episode started $\sim100\,\rm{Yr}$ and lasted up to $\sim20\,\rm{Yr}$ before the explosion of the progenitor star.
While a more accurate modeling of the observables is needed, a simple scenario might involve the presence of a H-rich companion as a possible source of the dense CSM needed to  produce the observables of \object.

\begin{acknowledgements}
The Oskar Klein Centre is funded by the Swedish Research Council.
Based on observations obtained with the Samuel Oschin Telescope 48-inch and the 60-inch Telescope at the Palomar Observatory as part of the Zwicky Transient Facility project. ZTF is supported by the National Science Foundation under Grant No. AST-1440341 and a collaboration including Caltech, IPAC, the Weizmann Institute for Science, the Oskar Klein Center at Stockholm University, the University of Maryland, the University of Washington, Deutsches Elektronen-Synchrotron and Humboldt University, Los Alamos National Laboratories, the TANGO Consortium of Taiwan, the University of Wisconsin at Milwaukee, and Lawrence Berkeley National Laboratories. Operations are conducted by COO, IPAC, and UW.
SED Machine is based upon work supported by the National Science Foundation under Grant No. 1106171
The data presented here were partly obtained with ALFOSC, which is provided by the Instituto de Astrofisica de Andalucia (IAA) under a joint agreement with the University of Copenhagen and NOTSA. \\
The Liverpool Telescope is operated on the island of La Palma by Liverpool John Moores University in the Spanish Observatorio del Roque de los Muchachos of the Instituto de Astrofisica de Canarias with financial support from the UK Science and Technology Facilities Council. \\
This research has made use of the NASA/IPAC Extragalactic Database (NED), which is funded by the National Aeronautics and Space Administration and operated by the California Institute of Technology. \\
This research has made use of the NASA/IPAC Infrared Science Archive, which is funded by the National Aeronautics and Space Administration and operated by the California Institute of Technology. \\
This work was supported by the GROWTH project funded by the National Science Foundation under Grant No 1545949. \\
Funding for the Sloan Digital Sky Survey (SDSS) has been provided by the Alfred P. Sloan Foundation, the Participating Institutions, the National Aeronautics and Space Administration, the National Science Foundation, the U.S. Department of Energy, the Japanese Monbukagakusho, and the Max Planck Society.
The SDSS Web site is \url{http://www.sdss.org/}. \\
The SDSS is managed by the Astrophysical Research Consortium (ARC) for the Participating Institutions. The Participating Institutions are The University of Chicago, Fermilab, the Institute for Advanced Study, the Japan Participation Group, The Johns Hopkins University, Los Alamos National Laboratory, the Max-Planck-Institute for Astronomy (MPIA), the Max-Planck-Institute for Astrophysics (MPA), New Mexico State University, University of Pittsburgh, Princeton University, the United States Naval Observatory, and the University of Washington. \\
The Pan-STARRS1 Surveys (PS1) and the PS1 public science archive have been made possible through contributions by the Institute for Astronomy, the University of Hawaii, the Pan-STARRS Project Office, the Max-Planck Society and its participating institutes, the Max Planck Institute for Astronomy, Heidelberg and the Max Planck Institute for Extraterrestrial Physics, Garching, The Johns Hopkins University, Durham University, the University of Edinburgh, the Queen's University Belfast, the Harvard-Smithsonian Center for Astrophysics, the Las Cumbres Observatory Global Telescope Network Incorporated, the National Central University of Taiwan, the Space Telescope Science Institute, the National Aeronautics and Space Administration under Grant No. NNX08AR22G issued through the Planetary Science Division of the NASA Science Mission Directorate, the National Science Foundation Grant No. AST-1238877, the University of Maryland, Eotvos Lorand University (ELTE), the Los Alamos National Laboratory, and the Gordon and Betty Moore Foundation. \\
L.T. acknowledges support from MIUR (PRIN 2017 grant 20179ZF5KS). \\
Y.-L.~K. has received funding from the European Research Council (ERC) under the European Unions Horizon 2020 research and innovation program (grant agreement No. 759194 USNAC). \\
This research has made use of the SVO Filter Profile Service (\url{http://svo2.cab.inta-csic.es/theory/fps/}) supported from the Spanish MINECO through grant AYA2017-84089 \\
{\sc iraf} is distributed by the National Optical Astronomy Observatory, which is operated by the Association of Universities for Research in Astronomy (AURA) under a cooperative agreement with the National Science Foundation. \\
{\sc SNOoPy} is a package for SN photometry using PSF fitting and/or template subtraction developed by E.~Cappellaro. A package description can be found at \url{http://sngroup.oapd.inaf.it/snoopy.html}. \\
{\sc foscgui} is a graphic user interface aimed at extracting SN spectroscopy and photometry obtained with FOSC-like instruments. It was developed by E.~Cappellaro. A package description can be found at \url{http://sngroup.oapd.inaf.it/foscgui.html}.

\end{acknowledgements}

\end{document}